*Title:* A process-independent explanation for the general form of Taylor's Law

*Article type:* Article

*Author affiliation:* Xiao Xiao[1, 2, 3, *], Kenneth J. Locey[4], and Ethan P. White[1, 2]

[1]Department of Biology, Utah State University, 5305 Old Main Hill, Logan, UT 84322-5305

[2]Ecology Center, Utah State University, 5205 Old Main Hill, Logan, UT 84322-5205

[3]Current affiliation: Mitchell Center for Sustainability Solutions, 5710 Norman Smith Hall, University of Maine, Orono, ME 04469

[4]Department of Biology, Indiana University, 1001 East 3rd Street, Bloomington, IN 47405

[*]Corresponding author

*Email addresses:* xiao@weecology.org; ken@weecology.org; ethan@weecology.org

**Abstract**

Taylor's Law (TL) describes the scaling relationship between the mean and variance of populations as a power-law. TL is widely observed in ecological systems across space and time with exponents varying largely between 1 and 2. Many ecological explanations have been proposed for TL but it is also commonly observed outside ecology. We propose that TL arises from the constraining influence of two primary variables: the number of individuals and the number of censuses or sites. We show that most possible configurations of individuals among censuses or sites produce the power-law form of TL with exponents between 1 and 2. This "feasible set" approach suggests that TL is a statistical pattern driven by two constraints, providing an *a priori* explanation for this ubiquitous pattern. However, the exact form of any specific mean-variance relationship cannot be predicted in this way, i.e., this approach does a poor job of predicting variation in the exponent, suggesting that TL may still contain ecological information.

**Introduction**

One of the most basic goals of ecology is to understand how ecological systems change across scales. Scaling relationships such as Kleiber's Law (Kleiber 1932; Brown et al. 2004) and the species-area relationship (Rosenzweig 1995) have been intensively studied for decades and serve as both useful tools for extrapolation and empirical targets for ecological theories. One of the most general scaling relationships is Taylor's Law (TL; Taylor 1961). TL proposes that the relationship between the variance ($s^2$) and the mean density ($m$) of populations is a power-law, which can be expressed mathematically as $s^2 \sim am^b$ where $a$ and $b$ are constants. TL has been confirmed as an adequate description of population fluctuations both spatially (Taylor 1961; Taylor and Taylor 1977; Taylor et al. 1978; He and Gaston 2003; Kaltz et al. 2012) and temporally (Taylor and Woiwod 1980; Perry 1994; Kerkhoff and Ballantyne 2003). Spatial TLs describe how populations vary in spatial aggregation as a function of their average size, with the points being the mean and the variance of a set of spatially distinct locations within a population. Temporal TLs describe the relationship between the level of temporal fluctuation of populations and their mean densities, with each point being the mean and variance of a single population time-series. The exponent $b$, which characterizes the curvature of the relationship, is generally bounded between 1 and 2 for both types of relationships (Taylor and Woiwod 1982). These patterns have been observed across thousands of studies from a diverse array of taxonomic groups (Taylor et al. 1978; Taylor and Woiwod 1980; Taylor et al. 1983), making Taylor's Law one of the most widely-documented patterns in ecology.

A number of different models based on distinct ecological processes have been proposed to explain Taylor's Law. Explanations for the spatial TL include density-dependent population growth (Perry 1994), density-independent population growth (Cohen et al. 2013), random walks of individuals in space (Hanski 1980), and simultaneous attraction and repulsion among conspecific individuals (Taylor 1981a; Taylor 1981b). Similarly, the temporal TL has been argued to arise from environmental and demographic stochasticity (Ballantyne 2005; Ballantyne and Kerkhoff 2007), interspecific competition (Kilpatrick and Ives 2003), and even sampling error (Kalyuzhny et al. 2014). However, similar power-law mean-variance relationships with exponents between 1 and 2 have recently been documented in a number of non-ecological systems, ranging from the distribution of genes on a chromosome (Kendal 2003) and the number of cells in individuals (Azevedo and Leroi 2001) to fluctuations in the stock market (Eisler and Kertész 2006) and traffic flow (de Menezes and Barabási 2004) (see Eisler et al. 2008 for a review). This suggests that the mechanism underlying the pattern may not be specific to ecological systems but may instead be purely statistical.

An alternative explanation to process-based models is that TL describes the mean-variance scaling relationship of most possible states of a system where individuals are divided among groups (such as censuses or plots). This constraint-based view has been explored for another common pattern, the species-abundance distribution, using the set of all possible configurations of the pattern given some constraints (the feasible set; Locey and White 2013) to determine if the general shape of the pattern simply reflects that of the majority of possible outcomes. This kind of reasoning would suggest that TL is not generated by any particular set of processes, but emerges because many different combinations of processes result in the same general pattern (Harte 2011; White et al. 2012; Frank 2014).

To investigate this possibility for TL, we begin by recognizing that the form of TL is necessarily influenced by two values: the total number (quantity) of individuals ($Q$), and the number of groups ($N$; e.g., plots in spatial TL, censuses in temporal TL) among which those individuals are distributed, for each point in the mean-variance relationship. Given these constraints we ask whether most possible mean-variance relationships take a roughly power-law form. We adopt the concept of the feasible set (Haegeman and Loreau 2008; Locey and White 2013), which provides a general context under which the observed patterns can be examined. By comparing the empirical TLs to mean-variance relationships created from randomly sampling the set of all possible relationships constrained by $Q$ and $N$, we find that both the form of the power-law and the exponent $b$ between 1 and 2 are expected to occur for most possible arrangements of the system; though the exact shape of each individual relationship cannot be accurately characterized without additional information.

**Methods**

1. The feasible set approach

The feasible set is the set of all possible configurations of a system (Haegeman and Loreau 2008; Locey and White 2013). This concept can be applied with different sets of constraints and configurations (e.g. ordered or unordered vectors of labeled individuals, ordered or unordered vectors of abundances, etc.). Since each mean-variance pair $(m_i, s_i^2)$ in a TL relationship results from the variation of abundances among $N_i$ groups, the sum of which is $Q_i$ and the mean of which is $m_i = Q_i / N_i$, we adopted $(Q_i, N_i)$ as a minimal set of constraints. The value of the unbiased sample variance $s_i^2$ depends on how the abundances vary among the groups, with a minimal value of 0 (assuming that $Q_i$ can be equally divided among $N_i$ groups) and a maximal value of $Q_i^2 / N_i$ (where the abundance is $N_i$ in one group and zero everywhere else).

We used two types of combinatorial objects, integer partitions and integer compositions, as configurations for feasible sets that may give rise to TL. Integer partitions are lists of unordered non-negative integers, usually written in non-increasing order. In contrast, integer compositions are ordered vectors of non-negative integers (Bona 2006; Severs and White 2010). For example, the feasible set of integer partitions for $Q_i = 4$ and $N_i = 2$ is: (4, 0), (3, 1), (2, 2). In this way, differently ordered configurations with the same integer values, e.g., (4, 0) and (0, 4), represent the same integer partition (4, 0) but different compositions. The difference between partitions and compositions is analogous to the difference between two well-known combinatorial objects, combinations and permutations, where combinations are unordered sets of numbers or objects that can be reordered into potentially many permutations. See Table 1 for definitions of technical terms.

The feasible set for TL may greatly differ depending on whether the order of the groups is considered (compositions) or not (partitions). Using integer compositions as configurations to explore TL is equivalent to shifting the weights of the partitions, i.e., making some configurations more likely to arise due to the differences in the number of compositions each partition has. In the above example of $Q_i = 4$ and $N_i = 2$, the partition (4, 0) has the same frequency as (2, 2) in the feasible set of partitions, but twice the frequency in the feasible set of compositions because there is no other way to reorder (2, 2). In fact, for a given ($Q_i$, $N_i$) pair, a partition with all unique numbers will always be associated with more compositions than a partition with repeated numbers, and hence greater representation in the feasible set of compositions. We examined the application of both partitions and compositions to TL because there is no *a priori* reason to favor one over the other, and because using both would provide a more complete context for understanding the constraining influence of $Q_i$ and $N_i$.

2. Data

To explore whether the feasible set could generate realistic empirical patterns, we compiled TL relationships from the literature by surveying all papers that cited Taylor (1961) on Google Scholar to which we had access. We collected all empirical relationships that directly reported the values of ($m_i$, $s_i^2$) and ($Q_i$, $N_i$) pairs, or contained enough information for these values to be calculated. Due to the limited number of available non-ecological TL relationships, we focused exclusively on ecological TLs describing spatial or temporal fluctuations of population abundances. This approach to data compilation resulted in an imbalance between the spatial and the temporal TLs, with 90 spatial relationships and only 4 temporal ones. To offset this imbalance, we added a compilation of community-level time-series data (Yenni 2013) to our analysis, increasing the number of temporal TLs to 113.

All relationships were put through additional screening before analysis. We removed ($m_i$, $s_i^2$) pairs where the corresponding $N_i$ was less than 3 to ensure that the variance $s_i^2$ was properly defined among at least 3 numbers, as well as pairs where the corresponding $Q_i$ was less than 5 so that the shape of TL would not be distorted by these zero-inflated, over-constrained configurations (Taylor and Woiwod 1982). We then excluded those relationships with less than 5 pairs of ($m_i$, $s_i^2$) remaining, leaving 73 spatial TLs and 106 temporal TLs. Due to the computationally intensive nature of the algorithm of generating partitions (Locey and White 2013), we further dropped TL relationships that contained any ($Q_i$, $N_i$) pairs that individually would take 2 hours or more to analyze (see 3. Analyses below). Overall our study encompassed 115 TL relationships, where 45 were spatial and 70 were temporal (see **Appendix A** for a full list of empirical relationships and their characteristics).

3. Analyses

In order to examine whether most possible mean-variance relationships exhibit TL-like behavior, we need to define a range of scenarios to explore. To avoid selecting arbitrary, and potentially unrealistic, values of $Q_i$ and $N_i$ (and the distributions of those values within individual datasets), we followed Xiao et al. (2011) and used empirical data to define the range of values explored. For each empirical mean-variance relationship, we generated its artificial counterparts from the feasible set using the following the steps:

1. We constructed the feasible set of partitions and the feasible set of compositions for each pair of ($Q_i$, $N_i$) in the empirical relationship. Given the large size of the feasible set for large values of ($Q_i$, $N_i$) (Locey and White 2013), we drew 1000 random configurations from the feasible set for each ($Q_i$, $N_i$) using the algorithms from Locey and McGlinn (2013). We confirmed that 1000 samples were sufficient in this context by testing sample sizes of 4000, which yielded equivalent results (**Appendix B**, Fig. B2).

2. We computed $s_{ij}^2$ for each sampled configuration, with $j$ ranging from 1 to 1000, leading to 1000 $s_{ij}^2$'s for each empirically observed $s_i^2$.

3. We built artificial mean-variance relationships by replacing $s_i^2$'s in the empirical relationship with $s_{ij}^2$'s that we obtained from the feasible set. Thus the first artificial relationship consisted of pairs of ($m_i$, $s_{i1}^2$), the second relationship consisted of pairs of ($m_i$, $s_{i2}^2$), etc., resulting in 1000 simulated mean-variance relationships, or 1000 arrangements of ($m_i$, $s_{ij}^2$) pairs, for each empirical relationship.

We examined the 95% (2.5% to 97.5%) quantiles of the 1000 $s_{ij}^2$'s for each ($Q_i$, $N_i$) pair to see if the variance for configurations within the feasible set was aggregated with most configurations having similar variance values, and if the corresponding empirical $s_i^2$ took a similar value. We then compared the mean-variance relationships obtained from sampling the feasible set to the empirical TL relationships with the same vectors of $Q_i$ and $N_i$. We fit power-law relationships to each empirical relationship and its 1000 corresponding simulated relationships using OLS regression on log-transformed data, which is the standard approach to assessing TL relationships. We compared the empirical and the simulated relationships based on the values of their slopes (an estimate of the exponent $b$), statistical significance, goodness-of-fit of the regression line (quantified with $R^2$), and deviation from the power-law (quantified as curvature on logarithmic scale, or the significance of a quadratic term fit to the relationship). All comparative analyses were conducted separately for samples from the feasible set of partitions and those from the feasible set of compositions.

**Results**

Fig. 1 depicts three empirical mean-variance relationships from our compilation together with simulated relationships from the feasible sets of partitions and of compositions to visually illustrate how the power-law allometries can be recovered by randomly sampling the feasible sets. Consistent with results from previous studies (e.g., Taylor and Woiwod 1982), most of the empirical datasets in our compilation have a significant power-law relationship between the mean and the variance, with little curvature, high $R^2$, and $b$ largely between 1 and 2 (Table 2). Mean-variance relationships simulated from the feasible set are well characterized by power-laws as well, with an average $R^2$ of 0.88 for relationships from the feasible set of partitions and 0.93 for those from the feasible set of compositions, though the latter show significant curvature (i.e., significant quadratic terms) in a non-negligible proportion of cases (27.8%, Table 2). The exponent $b$ estimated for the simulated relationships also largely falls between 1 and 2 (85.1% for partitions, 75.0% for compositions). Comparison between empirical TLs and those constructed from the feasible sets shows general agreement both in fit ($p$-values and $R^2$) and in the values of $b$ (Table 2, Fig. 2). Specifically, relationships constructed from the feasible set of partitions provide a surprisingly good match to empirical observations in the distribution of all statistics, while those from the feasible set of compositions exhibit deviations with higher proportions of significant curvature and exponents shifted towards higher values (Fig. 2).

On the other hand, comparison between individual empirical $s_i^2$'s and those calculated for random configurations in the feasible sets show that a high percentage of $s_i^2$'s fall outside of the 95% quantiles of the feasible set values for both spatial TLs (43.3% for partitions, 73.0% for compositions) and temporal TLs (26.5% for partitions, 41.9% for compositions) (Fig. 3, top panels). Examination of $s_{ij}^2$'s for individual $(Q_i, N_i)$ pairs shows that considerable variation exist for at least some combinations of $Q_i$ and $N_i$, leading to distributions of $s_{ij}^2$'s that are dispersed without a sharp peak where most configurations are aggregated (**Appendix B**, Fig. B1). This is particularly true for partitions. Consequently, the exponent $b$ estimated for individual empirical TLs falls outside of the 95% quantile of those generated from the feasible set in 40.0% of spatial TLs and 28.6% of temporal TLs for partitions (73.3% and 60.0% for compositions; Fig. 3, bottom panels), and the feasible set values of the exponent do not provide good predictions for the empirical values (coefficient of determination = -0.035 for partitions, -0.61 for compositions). These results suggest that neither feasible set approach is able to accurately characterize the variance of individual observations, or the exact shape of individual mean-variance relationships. While most possible mean-variance relationships are well characterized by power-laws, the precise form of the power-law for a particular set of $(Q_i, N_i)$ is not strongly constrained by the feasible set.

**Discussion**

Taylor's Law, or the power-law relationship between the mean and the variance of one or more populations, is a general pattern that has been widely observed in both ecological and non-ecological systems (Eisler et al. 2008). While numerous processes have been proposed as explanations for this pattern (e.g., Taylor 1981a; Kilpatrick and Ives 2003; Ballantyne and Kerkhoff 2007; Cohen et al. 2013), our study shows that TL can arise from numerical constraints on the system without explicitly or implicitly invoking processes. By sampling the set of all possible mean-variance scaling relationships we find that the form of the power-law relationship is exceptionally robust (Table 2, Fig. 2) with distributions of the exponent $b$ closely matching those estimated from the empirical relationships (Fig. 2D), despite the relatively high observable variation for variance values among the configurations (Fig 3, Fig. B1). Thus the feasible set provides a general explanation for the ubiquity of TL in nature and helps explain why the pattern can be produced by models based on different underlying processes.

Our study is not the first to suggest that TL may be statistical in nature. By applying the Maximum Entropy Principle, Fronczak and Fronczak (2010) proposed that TL is the most likely macrostate associated with the largest number of microstates. However, their approach was later criticized for its reliance on physical quantities such as free energy and an external field, which lack analogues in biological systems (Kendal and Jørgensen 2011). Kendal and Jørgensen (2011) suggested instead that TL is associated with the Tweedie distribution family (Tweedie 1984), which by definition are characterized by a strict power-law relationship between the variance and the mean. While it has been argued that many statistical systems converge to distributions in the Tweedie family as limiting cases (Tweedie convergence theorem; Jørgensen et al. 1994), it is not clear how such convergence is achieved in nature.

Our approach shows that most of the possible configurations for dividing a given total among a number of groups will result in a mean-variance scaling relationship consistent with Taylor's Law. This approach is unique in that it is built upon the concept of the feasible set, which can be unambiguously defined and applied to any system. With the feasible set approach,

Locey and White (2013) showed that the majority of the configurations constrained by total abundance $N$ and total species richness $S$ conform to a hollow-curve similar in shape to empirically observed species-abundance distributions (SAD). Our study shows similar results for a second general pattern, TL, illustrating again that patterns can arise simply as the aggregated central tendency within a feasible set. This further emphasizes the importance of examining empirical patterns in the context of the forms that the pattern is capable of taking.

The feasible set approach is part of a more general framework to understanding macroecological patterns that has been emerging over the last decade. These approaches, which also include applications of the Maximum Entropy Principle, propose that some empirical patterns are emergent statistical properties constrained by numerical inputs (state variables), while the processes operate only indirectly through their effects on constraints formed by the state variables (Harte 2011; Supp et al. 2012; Locey and White 2013; Frank 2014). The constraints, which are usually descriptive statistics of the system (i.e., state variables or moments of the distributions), either strictly limit the possible configurations that a system can take ("hard" constraints; e.g., there are a limited number of ways to allocate 100 individuals into 5 groups) or limit the expected characteristics of the system ("soft" constraints"; e.g., 20 individuals are expected on average for each group but the observed values may vary; see Haegeman and Etienne 2010). In our study the system is hard-constrained by the total number of individuals observed ($Q_i$) and the number of groups they belong to ($N_i$), which forces the mean of configurations in the feasible set to match the mean of empirical data. The power-law relationship between the mean and the variance then arises for most possible arrangements of ($m_i$, $s_i^2$) pairs generated by randomly sampling the feasible set.

While the application of the feasible set approach is independent of assumptions about processes, the distribution of possible states of the system (in our case, variance) may shift with different definitions of what constitutes a unique configuration. This can be seen by the differences between the feasible set of partitions and the feasible set of compositions (Figs 2, 3). Partitions describe how a total can be divided among groups with no information on the identity of either the individuals or the groups. Applied to ecological populations, it is equivalent to simultaneously assigning quantities of individuals at random to a number of plots or censuses, focusing exclusively on the possible forms of the pattern. This deliberately ignores how the relative likelihood of a particular pattern could vary depending on specific null modeling decisions related to the identity of the individuals or the plots or censuses that they belong to. As such, this is not a null model in the traditional sense, where inference is drawn for a specific process by comparing the empirical data with the null which does not include the process. Instead, this approach tells us about the possible patterns that can occur, and therefore provides context for whether it will be possible to associate any specific process with the pattern in question (Locey and White 2013).

While partitions were the only configuration used for understanding the possible forms of the species-abundance distribution (Locey and White 2013), we also considered integer compositions in this study. This configuration is the same as simultaneously choosing quantities of individuals at random to be assigned to specific sites or surveys. Unlike partitions, there may be several compositions that result in the same state of the pattern, i.e., multiple compositions can be associated with the same partition and thus have the same value of variance. Compositions are also closer to a classic null modeling approach because it is possible to envision population level processes, such as competition, that are being excluded in this approach to distributing populations of individuals. There are two other configurations that we did not consider, both of which focus on the distribution of individual organisms. One is the equivalent of randomly selecting the location of each of a set of known individuals among sites or censuses (known as surjections in combinatorics); e.g., the composition (1, 1) can be expanded into two configurations (A, B) and (B, A), where A and B represent labeled individuals. This is how most traditional null modeling is conducted in ecology. The other is the case where elements of the configuration are labeled but the identities of their bins are not accounted for; e.g., knowing that individuals A and B are in the same plot but not which plot they are located in. These two configurations are excluded because we are interested in patterns that are related to how entire populations of individuals are distributed across space and through time, not the specific location of individual organisms.

While there is a reasonable justification for choosing to focus on how populations are distributed rather than how individuals are distributed, both partition- and composition-based approaches satisfy this consideration. In equivalent discussions on how to set up maximum entropy based models, it has been argued that these choices must be made on the basis of comparisons to empirical data (Haegeman and Etienne 2010). Our study shows that the feasible set of partitions provides a more adequate characterization for our compilation of ecological TL relationships both in the fit of the power-law form (Table 2) and in the distribution of the exponents (Fig. 2D), however it remains to be seen if this statement holds more broadly, especially in non-ecological systems.

Our broad-stroke explanation for Taylor's Law based on the majority of possible outcomes from the feasible set does not preclude the influence of other processes and constraints. In fact, our research shows that the exponent $b$ for a particular relationship cannot be accurately estimated by merely examining the central tendency of the feasible sets defined by $(Q_i, N_i)$ alone

(Fig. 3). Many previous studies have shown that *b* is often characteristic of a given system which may reflect underlying ecological properties. In spatial TLs *b* is strongly tied to spatial aggregation (He and Gaston 2003), while different values of *b* in temporal TLs have been attributed to the interplay between synchronized versus independent reproduction (Ballantyne and Kerkhoff 2007), binomial versus proportional sampling error (Kalyuzhny et al. 2014), and the strength of interspecific competition (Kilpatrick and Ives 2003). While our results suggest that both the power-law form of the relationships and the general bound on their exponents are likely to be statistical, failure of the feasible set in pinning down the specific values of *b* for individual relationships implies that the ecological significance of these relationships may lie in their detailed shapes, which can only be understood with more information of the systems that go beyond the constraints $Q_i$ and $N_i$.

**Acknowledgements**

We thank the authors of the papers and datasets included in our analysis for making their data publicly available, and Glenda Yenni for sharing with us the data she compiled for her own research. A. J. Kerkhoff and two anonymous reviewers have provided valuable comments which greatly improved the presentation of our work. This research was originally inspired by Joel Cohen's talk on Taylor's Law at INTECOL 2013 (London). It was supported by a CAREER award from the U.S. National Science Foundation to E. P. White (DEB-0953694). Data sources include the Desert Laboratory, supported by NSF grants DEB 9419905 (LTREB), DEB 0212782 (LTREB), and DEB 0717466 (LTREB); HBES LTER (with data on bird abundance at Hubbard Brook and Lepidoptera larvae abundance in northern hardwood forests, provided by Richard T. Holmes), a collaborative effort at the Hubbard Brook Experimental Forest operated and maintained by the USAD Forest Service, Northern Research Station, Newtown Square, PA; Jornada Basin LTER, supported by the NSF grant DEB-1235828; KNZ LTER, supported by the NSF Long Term Ecological Research Program; Luquillo LTER, supported by grants BSR-8811902, DEB 9411973, DEB 0080538, DEB 0218039, DEB 0620910 and DEB 0963447 from NSF to the Institute for Tropical Ecosystem Studies, University of Puerto Rico, and to the International Institute of Tropical Forestry USA Forest Service, with additional support from the U.S. Forest Service (Dept. of Agriculture) and the University of Puerto Rico; SANParks; Sevilleta LTER, supported by the National Science Foundation Long Term Ecological Research program with NSF grant numbers BSR 88-11906, DEB 9411976, DEB 0080529 and DEB 0217774; and Virginia Coast Reserve LTER, supported by NSF Grants BSR-8702333-06, DEB 9211772, DEB 9411974, DEB 0080381 and DEB 0621014. Results presented in this study are a modified version and not the original data and documentation distributed by the data owners.

**Table 1.** Definition of terminology used in this paper, where $m_i$ is the mean and $s_i^2$ is the variance for a $(Q_i, N_i)$ pair.

| Arrangement | A mean-variance relationship with all its $(m_i, s_i^2)$ pairs. |
|---|---|
| Configuration | The organization of a system's elements. Here, configurations refer to the organization of $N_i$ non-negative integers that sum to $Q_i$, with or without respect to order. |
| Feasible set | The complete set of all possible configurations of a system, given a set of constraints. |
| Composition | A unique list of ordered, non-negative integers. |
| Partition | A unique list of unordered, non-negative integers. |

**Table 2.** Summary of the mean-variance relationships constructed from sampling the feasible sets of partitions or compositions, as well as empirical spatial and temporal TL relationships.

|  | Significance $\alpha = 0.05$ | Curvature $\alpha = 0.05$ | Average $R^2$ | Proportion of $b$ between 1 and 2 |
|---|---|---|---|---|
| Partitions | 95.2% | 8.49% | 0.88 | 85.1% |
| Compositions | 98.0% | 27.8% | 0.93 | 75.0% |
| Spatial TLs | 86.7% | 11.1% | 0.80 | 77.8% |
| Temporal TLs | 97.1% | 4.29% | 0.86 | 88.6% |

**Figure Legends**

**Figure 1.** Three empirical mean-variance relationships (one in each column) versus one realization of their corresponding simulated relationships from the feasible set of partitions (top row) or compositions (bottom row). The fitted lines are obtained with OLS regression. The inset in each subplot represents the density distribution of the exponent $b$ estimated from all 1000 simulated relationships (green or magenta curve), versus $b$ for the empirical relationship (black vertical line). Note that the simulated relationships largely retain the power-law allometry; however their $b$ values can span a relatively broad range, the mean of which may or may not match the empirical $b$.

**Figure 2.** Density distribution of (A) p-values for the exponent $b$, (B) p-values for the quadratic term, (C) $R^2$ of the power-law relationship on logarithmic scale, and (D) values of the exponent $b$. Empirical results from the spatial TLs and the temporal TLs are not qualitatively different and are thus pooled. The dashed vertical lines in (A) and (B) represent the significance level of 0.05.

**Figure 3.** Comparison between the empirical values and their counterparts from the feasible sets of partitions (left) and compositions (right). Top panels show the results for variance ($s^2$), while bottom panels show the results for $b$. Vertical bars represent the 95% quantile for each point. For visualization purposes, values obtained from the feasible sets are standardized with respect to empirical values, either by division (for $s^2$) or by substraction (for $b$). Each subplot is divided into three sections by vertical dashed lines – points where the 95% quantile from the feasible set is below the empirical value (left), where it covers the empirical value (middle), and where it is above the empirical value (right). The horizontal line is where the mean value from the feasible set equals the empirical value. Results from spatial and temporal relationships are lumped.

**Figure 1.**

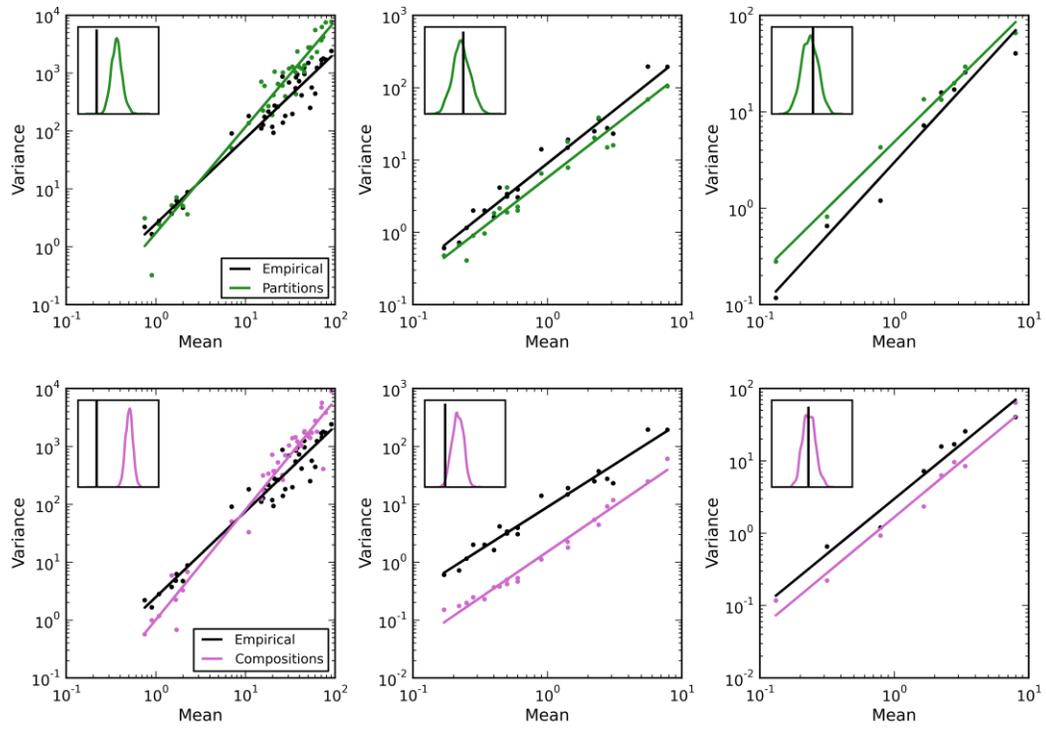

**Figure 2.**

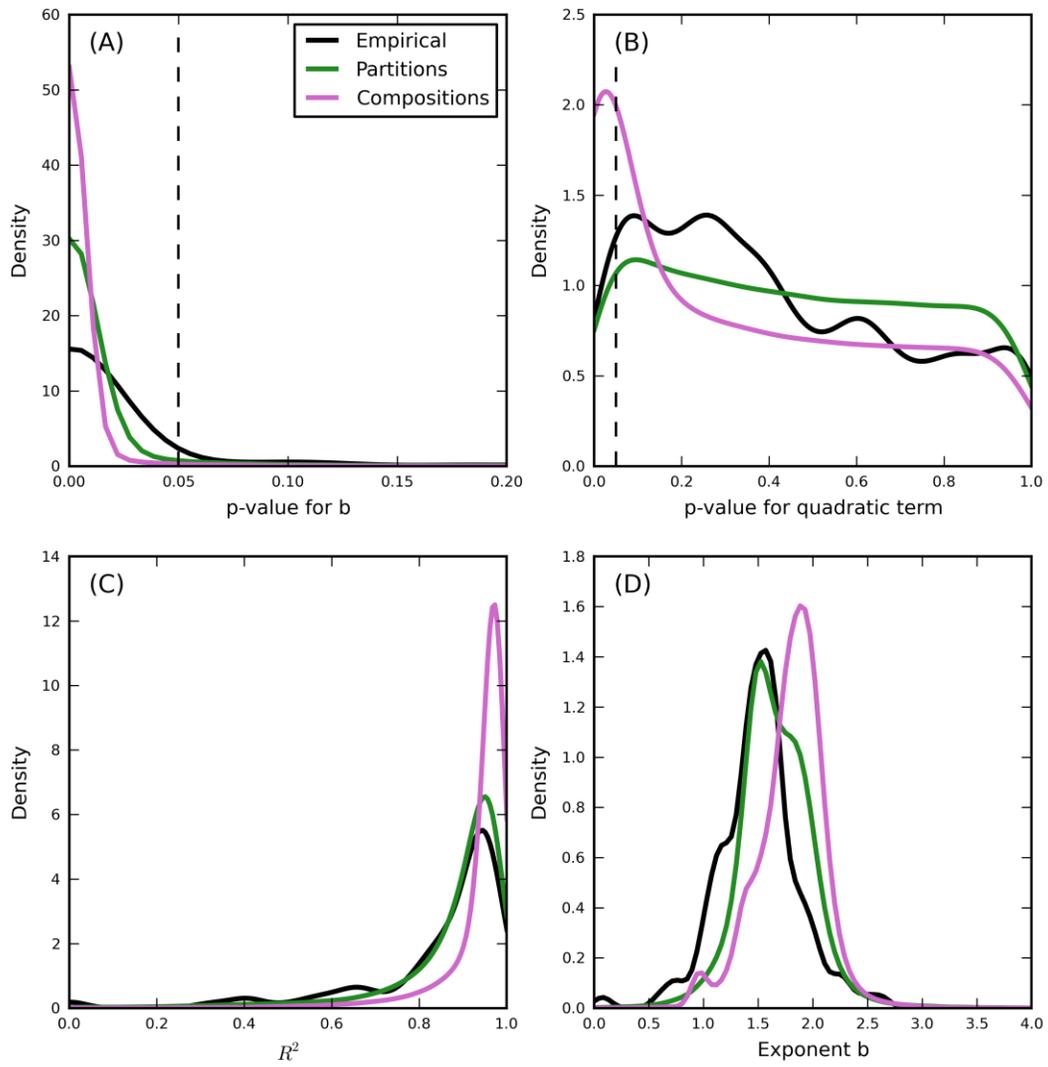

**Figure 3.**

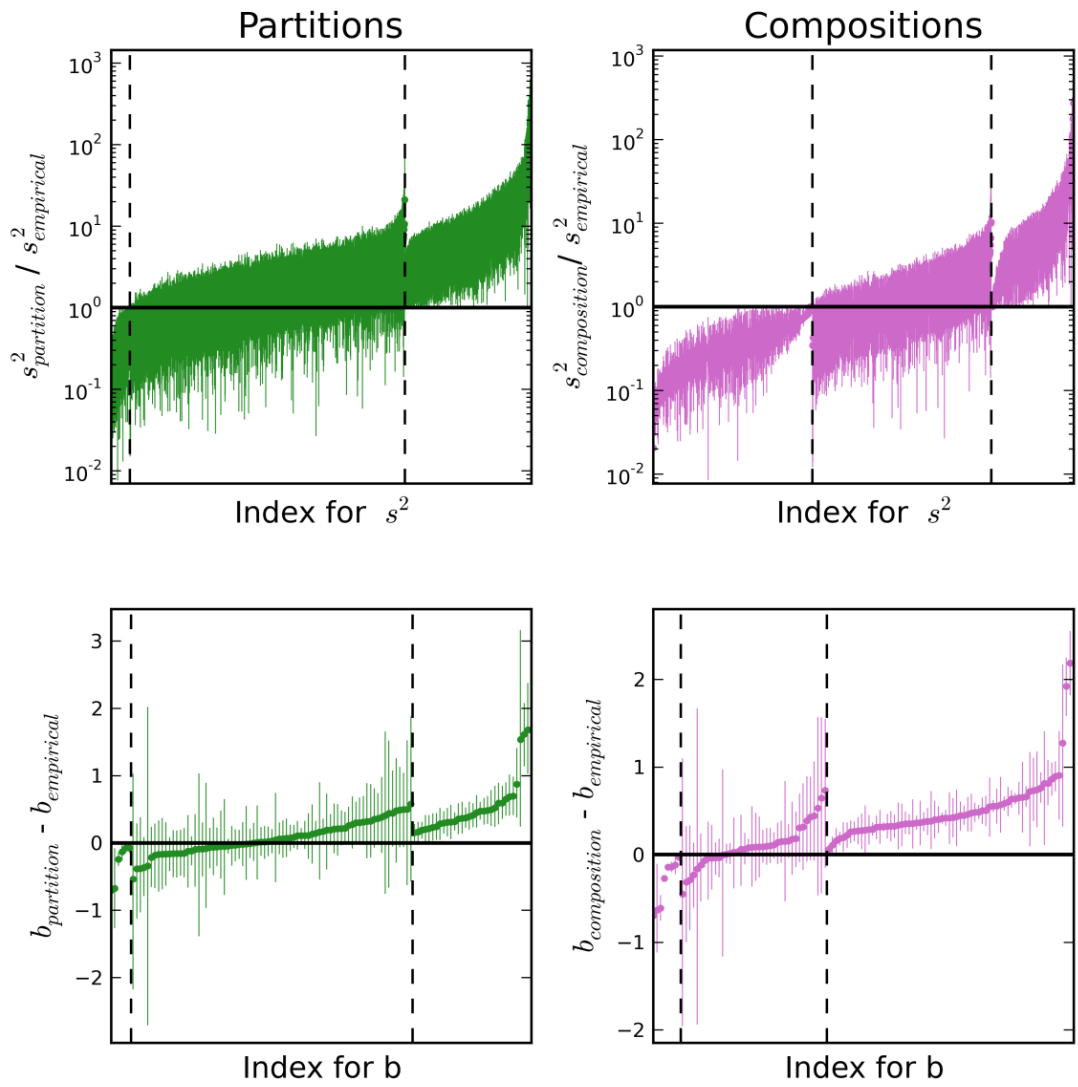

# Appendix A. Information on Compiled Datasets

**Table A1.** Summary of datasets.

| Study ID | Taxon | Type | Number of Data Sets | Number of Data Sets Included | Reference |
|---|---|---|---|---|---|
| 1 | fish | spatial | 1 | 1 | Stanfield et al. 2013 |
| 2 | bacteria | spatial | 16 | 0 | Kaltz et al. 2012 |
| 3 | arthropod | temporal | 2 | 2 | Karban et al. 2012 |
| 4 | arthropod | spatial | 2 | 2 | Hui et al. 2012 |
| 5 | arthropod | spatial | 3 | 3 | Thein and Singh 2011 |
| 7 | arthropod | spatial | 4 | 4 | Costa et al. 2010 |
| 8 | nematode | spatial | 2 | 0 | Aminayanaba 2010 |
| 10 | arthropod | spatial | 1 | 1 | Lessio and Alma 2006 |
| 11 | fungi | spatial | 4 | 0 | Sallam et al. 2007 |
| 12 | arthropod | spatial | 2 | 0 | Nachman 2006 |
| 13 | mammal | spatial | 1 | 0 | McMahon et al. 2005 |
| 14 | arthropod | spatial | 1 | 1 | Sileshi and Magongoya 2004 |
| 15 | invertebrate | spatial | 1 | 1 | Clarke et al. 2002 |
| 16 | arthropod | spatial | 1 | 1 | Silva et al. 2003 |
| 17 | arthropod | spatial | 8 | 0 | Parker et al. 2002 |
| 18 | annelid | spatial | 11 | 6 | Jiménez et al. 2001 |

| 19 | arthropod | spatial | 2 | 2 | Floater 2001 |
| --- | --- | --- | --- | --- | --- |
| 20 | arthropod | spatial | 1 | 1 | Schexnayder et al. 2001 |
| 21 | mollusc | spatial | 2 | 2 | Eleutheriadis and Lazaridou-Dimitriadou 2001 |
| 23 | mollusc | spatial | 1 | 1 | Babineau 2000 |
| 24 | mollusc | spatial | 1 | 1 | Staikou 1998 |
| 25 | arthropod | spatial | 1 | 1 | Pahl 1969 |
| 26 | mollusc | spatial | 1 | 1 | Todd 1978 |
| 27 | protist | spatial | 4 | 4 | Buzas 1970 |
| 28 | plant | spatial | 3 | 3 | Crawley and Weiner 1991 |
| 29 | fish | spatial | 1 | 0 | Van Damme and Hamerlynck 1992 |
| 30 | nematode | spatial | 2 | 0 | Warren and Linit 1992 |
| 31 | arthropod | spatial | 5 | 3 | Rosewell et al. 1990 |
| 32 | arthropod | temporal | 2 | 0 | Samways 1990 |
| 33 | Echinorhynchidae | spatial | 1 | 1 | Brattery 1986 |
| 34 | nematode | spatial | 4 | 2 | Wheeler et al. 1987 |
| 35 | arthropod | spatial | 1 | 1 | Purrington et al. 1989 |
| 36 | invertebrate | spatial | 2 | 2 | He and Gaston 2003 |
| 37 | bird | temporal | 1 | 1 | Dickson et al. 1993 |
| 38 | bird | temporal | 1 | 1 | Gaston and Blackburn 2000 |
| 39 | bird | temporal | 4 | 4 | Holmes et al. 2012 |
| 40 | bird | temporal | 1 | 0 | Sandercock 2009 |
| 41 | bird | temporal | 1 | 1 | Waide 2012 |
| 43 | bird | temporal | 1 | 1 | Vickery and Nudds 1984 |
| 44 | bird | temporal | 1 | 1 | Williamson 1983 |
| 45 | herp | temporal | 3 | 3 | How 1998; Thompson and Thompson 2005; Wilgers et al. 2006 |
| 46 | arthropod | temporal | 3 | 3 | Anderson 2003 |
| 47 | arthropod | temporal | 4 | 4 | Holmes 1997 |
| 48 | mollusc | temporal | 4 | 0 | Willig and Bloch 2004 |
| 49 | arthropod | temporal | 37 | 23 | Pollard et al. 1986 |
| 50 | mammal | temporal | 1 | 0 | Grant 1976 |
| 51 | mammal | temporal | 1 | 0 | Bestelmeyer 2007 |
| 52 | mammal | temporal | 16 | 15 | Kaufman 2010 |
| 53 | mammal | temporal | 1 | 0 | Merritt 1999 |
| 54 | mammal | temporal | 9 | 7 | Friggens 2008 |
| 55 | mammal | temporal | 7 | 2 | Stapp 2006 |
| 56 | mammal | temporal | 8 | 2 | SANParks 1989; SANParks 1997; SANParks 2004; SANParks 2009 |
| 57 | plant | temporal | 1 | 0 | Venable 2008 |
| 58 | plant | temporal | 1 | 0 | Adler et al. 2007 |
| 59 | plant | temporal | 1 | 0 | Clark and Clark 2006 |

| 60 | plant | temporal | 1 | 0 | Zachmann et al. 2010 |
| 62 | mammal | temporal | 1 | 0 | Ernest et al. 2009 |

# Appendix B. Additional Figures

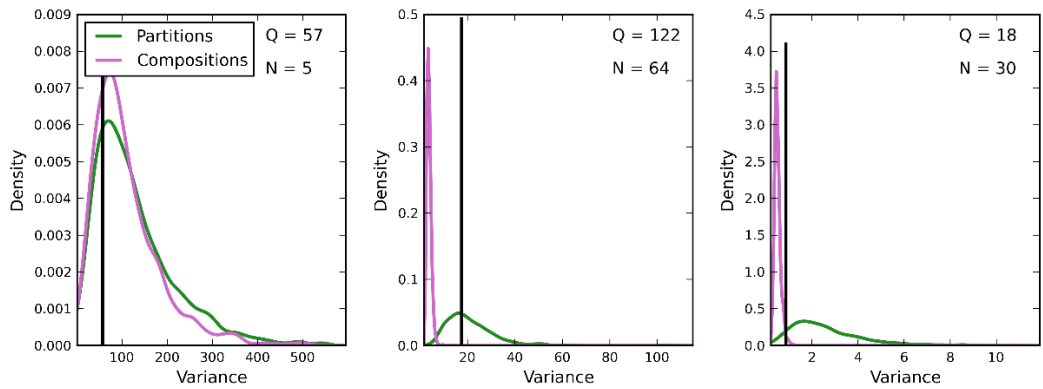

**Figure B1.** Examples showing the distribution of the variance calculated for configurations sampled from the feasible sets. The black vertical line in each subplot represents the empirical variance of the data from which the feasible sets are constructed.

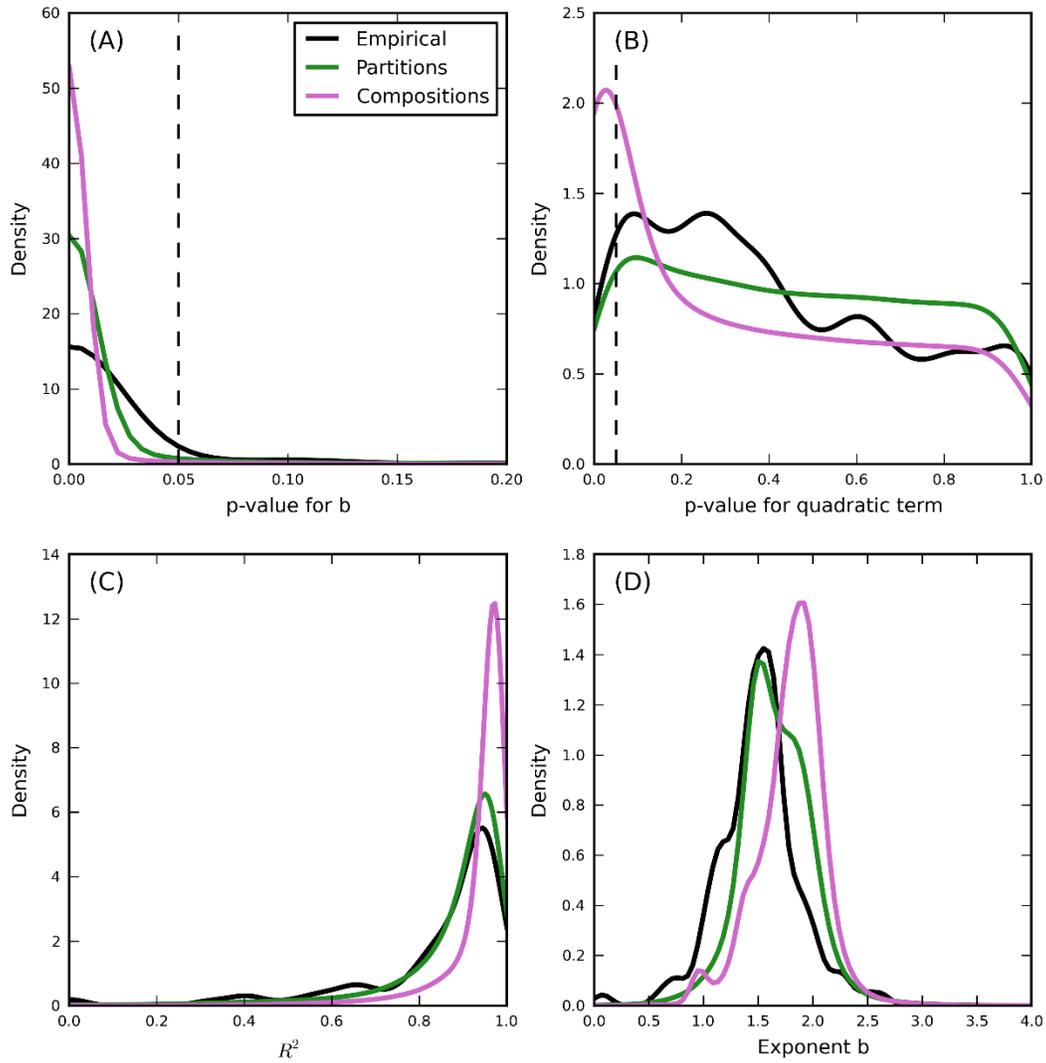

**Figure B2.** The equivalent of Fig. 2 in the main body of the paper, with the analyses conducted using 4000 (instead of 1000) simulated configurations for each ($Q_i$, $N_i$) pair. The two figures are almost identical, which implies that 1000 is a sufficient sample size to capture the variation within the feasible sets.